# Modeling and Analysis of Grid Tied Combined Ultracapacitor Fuel Cell for Renewable Application

[1]Webster Adepoju, *Student Member, IEEE*, [1]Indranil Bhattacharya, *Member, IEEE*
[2]Olufunke Mary Sanyaolu
[1]*Department of Electrical and Computer Engineering, Tennessee Technological University,*
[2]*Department of Material Science, University of Johannesburg*
Cookeville, 38505, TN, USA
woadepoju42@tntech.edu, ibhatacharya@tntech.edu

***Abstract*—In this manuscript, the performance of an ultracapacitor fuel cell in grid connected mode is investigated. Voltage regulation to the ultracapacitor was achieved with a three level bidirectional DC-DC converter while also achieving power flow from the grid to the ultra-capacitor via the bidirectional converter. The choice of a bidirectional three level converter for voltage regulation is based on its inherently high efficiency, low harmonic profile and compact size. Using the model equations of the converter and grid connected inverter derived using the switching function approach, the grid's direct and quadrature axes modulation indices, $M_d$ and $M_q$, respectively were simulated in Matlab for both lagging and leading power factors. Moreover, the values of $M_d$ and $M_q$ were exploited in a PLECS based simulation of the proposed model to determine the effect of power factor correction on the current and power injection to grid.**

***Index Terms*—Ultracapacitor, Inverter, Three level bidirectional DC-DC converter, Fuel cell, PLECS, Matlab**

## I. Introduction

In recent times, the quest for clean and emission-free energy generation has culminated in an increased penetration of renewable and distributed energy sources. Wind, tidal, solar, fuel cells and ultracapacitors are some of the renewable energy sources already explored to a commercially advanced stage. Conventional energy sources in the form of fossil fuels are inherently laden with dangerous pollutants, effectively threatening the the earth's eco-system [1]–[3]. While solar and wind power generation are affected by wind speed, insufficient solar irradiation, shading and temperature of the solar panel, fuel cells are preferred due their efficient and emission-free means of power generation. In addition, advances in fabrication technology coupled with increase in the storage capacity of batteries give it a competitive edge over other existing power generation sources, especially in the face of rising power demand both for domestic and industrial use. Further, the quest for reduction in carbon footprint coupled with its capability for high power density makes it a first in line industry-choice for vehicular applications, controlled electric drive and uninterruptible power supplies etc. In essence, they can be thought of as a battery bank or an ultracapacitor for charging and discharging operation [4]. It worth mentioning that even though fuel cells are popular for the aforementioned advantages, they are largely incapable of fast and rapid adjustment to electrical load transient. The above drawback can be attributed to their inherently slow electro-chemical and thermodynamic responses.

While ultra-capacitor have been widely researched and published in literature, there is no extant research that effectively describes the dynamic and steady state behavior of ultra-capacitor in a grid connected scenario. The main contribution of the work borders on the derivation of the dynamic and steady state mathematical model of an ultra-capacitor fuel cell in grid connected mode. In a system with high penetration of large loads, the adoption of DC-DC converter will be key to actualizing high power gain while increasing the conversion efficiency of the fuel cell. To this end, this manuscript presents a detailed dynamic and steady state mathematical models of a grid connected ultra-capacitor battery bank. In order to achieve this, a three-level bidirectional (TLB) buck-boost converter is interfaced in series with an ultra-capacitor fuel cells for efficient device operation. TLB has been proposed in different literature for various applications [1]–[4]. In [1], a TLB bidirectional converter was harnessed for operation of electric vehicles (EV) while the charging and discharging of ultra-capacitor based on a bidirectional TLB converter is embraced in [3], [4]. In addition, a novel dynamic and steady state model of fuel cells based on electrical circuits is proposed in [2]. In contrast with most classical DC-DC converters, a TLB converter has the advantage of dual operation and seamless transition between the buck and boost mode. This feature can be exploited for charging and discharging of a battery system. In line with the design proposed in this work, the ultra-capacitor fuel storage is charged and discharged depending on the current operating mode of the proposed TLB buck boost converter. Conventional DC-DC converters, including Cuk converter, boost converter [5], [6] etc. have inherent performance limitations, arising from high voltage and current ripples, coupled with large harmonic content resulting from switching losses. It is noteworthy that in an ultra-capacitor-TLC-inverter connection, the power flows from the grid to the converter and then to the ultra-capacitor/battery bank. Inherently, the charged ultra-capacitor serves as an energy repository for future use. Hence, in a grid-outage scenario, the inherent bidirectional operation of the converter comes in handy in changing the sequence of power flow from the battery bank to grid [7]–[9].

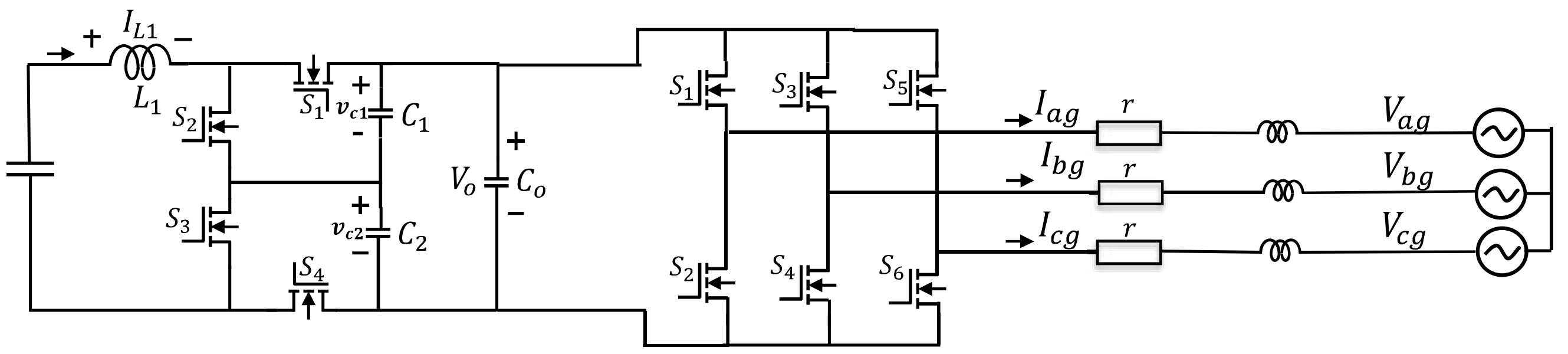

Fig. 1. Proposed Fuel Cell, Bidirectional Multilevel Buck Boost Converter and a Three Phase DC-AC Converter

The remainder of this manuscript is structured as follows: Section II presents a comprehensive model derivation of the TLC converter's voltage gain coupled with the stationary reference frame ($abc$) to synchronous reference frame ($dqo$) transformation of the model's DC-AC converter. Further, a comprehensive mathematical derivation, analysis and description of the dynamic and steady state equations of the proposed system is evaluated. This includes the $d$-axis and $q$-axis modulation indices and transformation from $abc$ to $qdo$ reference frames. The combined structure of the derived dynamic and steady state equations are scripted in Matlab to evaluate the modulation indices coupled with the real and active power injected into the grid. In section IV, a detailed analysis and discussion of the PLECS and Matlab based simulation results is presented. below.

TABLE I
LOOK UP TABLE OF THE CONVERTER SWITCHING STATE

| $S_1$ | $S_2$ | $S_3$ | $S_4$ | Switching State | Mode |
|---|---|---|---|---|---|
| 0 | 1 | 1 | 0 | $S_A = S_2S_3$ | Mode I |
| 0 | 1 | 0 | 1 | $S_B = S_2S_4$ | Mode II |
| 1 | 0 | 1 | 0 | $S_C = S_1S_3$ | Mode III |
| 1 | 0 | 0 | 1 | $S_D = S_1S_4$ | Mode IV |

## II. SYSTEM DESCRIPTION AND MODELING

### A. Three Level Buck-Boost Converter Modeling

In this section, a comprehensive mathematical derivation of the switch ON and OFF states of the proposed TLB converter is presented. Table I shows the various switching states of the converter for all the possible modes of operation. Further, the complete model representation of the proposed TLB is as shown in Fig. 1. The section to the left depicts a TLB buck boost converter. The four operational modes of the proposed system are analyzed below:

MODE I: $S_2$ and $S_3$ ON, $S_1$ and $S_4$ OFF

$$L\frac{dI}{dt} = V_s \tag{1}$$

$$C_1\frac{dV_{c1}}{dt} = -I_o \tag{2}$$

$$C_2\frac{dV_{c2}}{dt} = -I_o \tag{3}$$

MODE II: $S_2$ and $S_4$ ON, $S_1$ and $S_3$ OFF

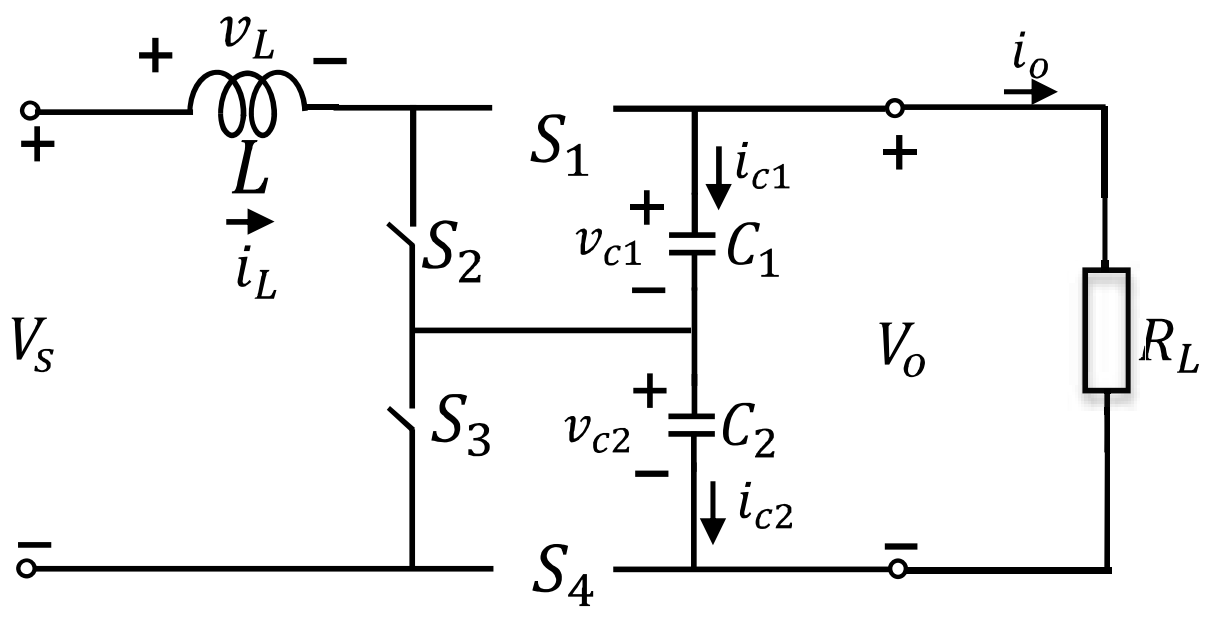

Fig. 2. Mode I

.

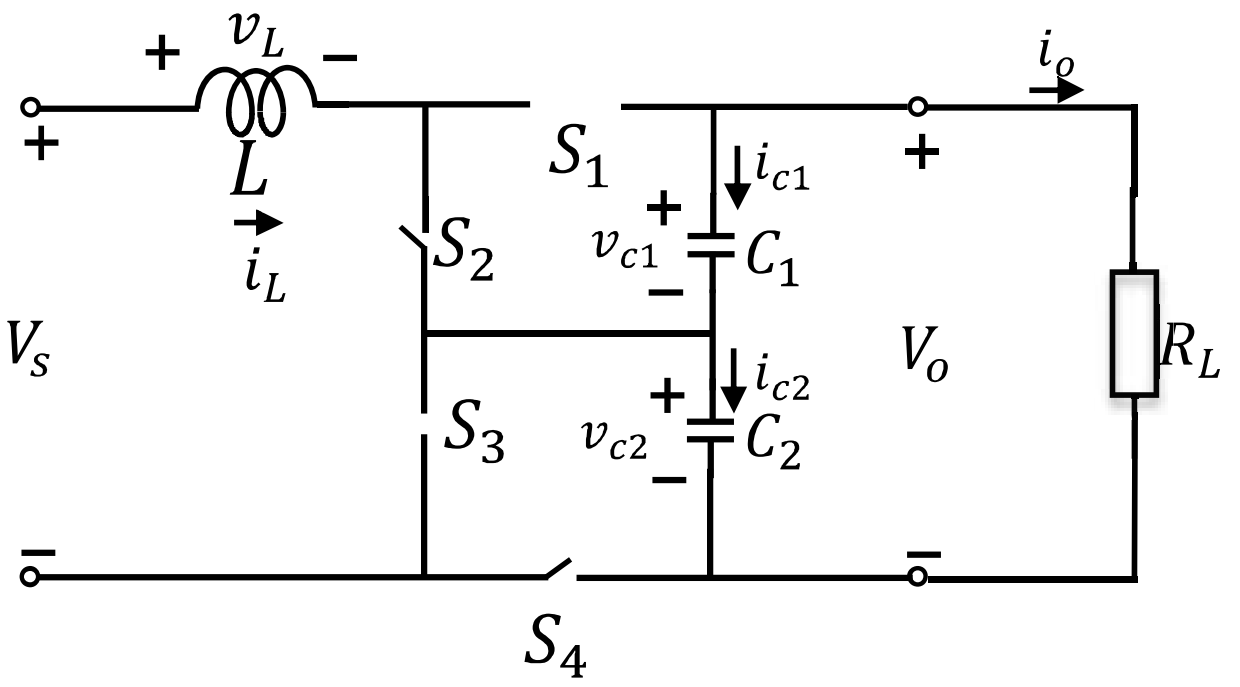

Fig. 3. Mode II

.

$$L\frac{dI}{dt} = V_s - V_{c2} \tag{4}$$

$$C_1\frac{dV_{c1}}{dt} = -I_o \tag{5}$$

$$C_2\frac{dV_{c2}}{dt} = I_L - I_o \tag{6}$$

MODE III: $S_1$ and $S_3$ ON, $S_2$ and $S_4$

$$L\frac{dI}{dt} = V_s - V_{c1} \tag{7}$$

$$C_1\frac{dV_{c1}}{dt} = I_L - I_o \tag{8}$$

$$C_2\frac{dV_{c2}}{dt} = -I_o \tag{9}$$

MODE IV: $S_1$ and $S_4$ ON, $S_2$ and $S_3$

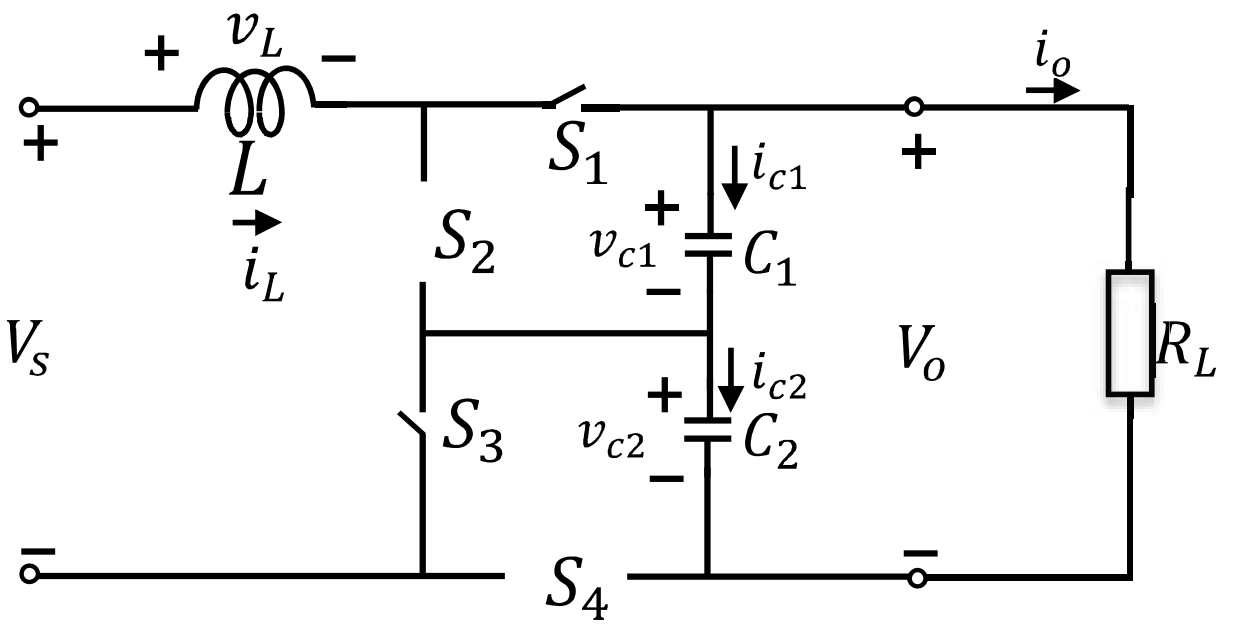

Fig. 4. Mode III

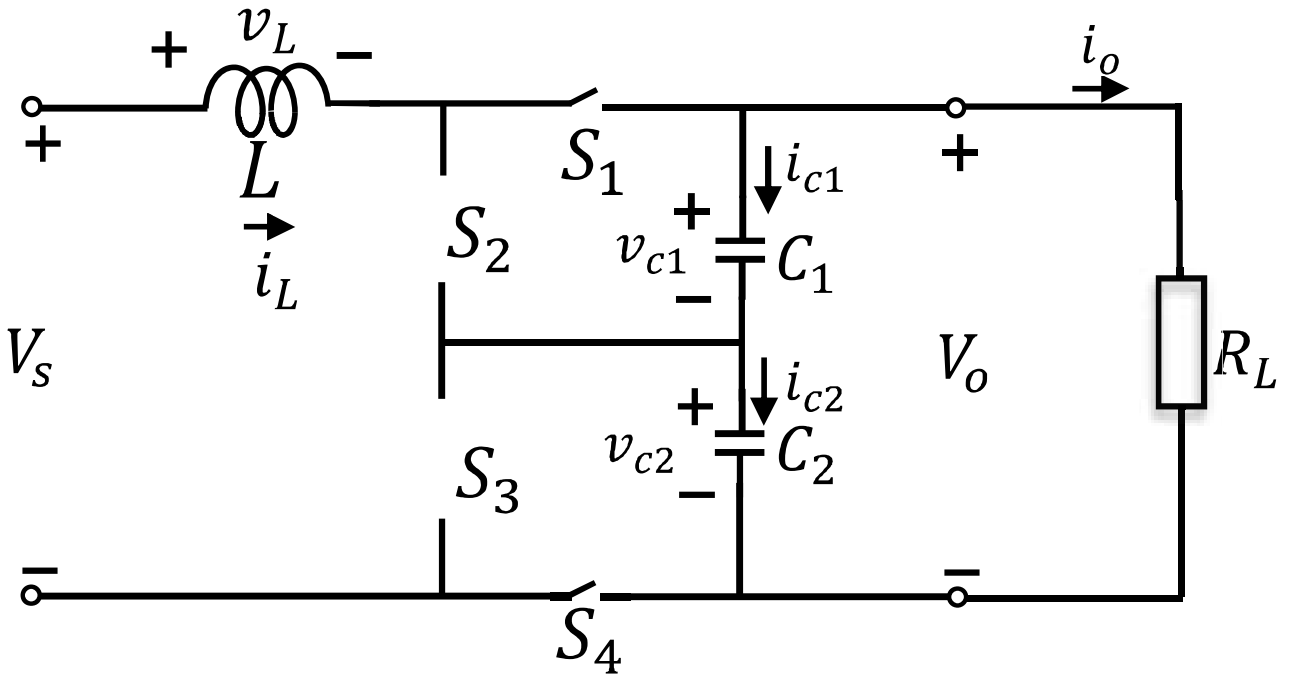

Fig. 5. Mode IV

$$L\frac{dI}{dt} = V_s - V_{c1} \tag{10}$$

$$C_1\frac{dV_{c1}}{dt} = I_L - I_o \tag{11}$$

$$C_2\frac{dV_{c2}}{dt} = -I_o \tag{12}$$

By applying the switching functions, the following equations are obtained:

$$(S_A + S_B + S_C + S_D)L\frac{dI}{dt} = S_AV_s + S_A(V_s - V_{c2}) + S_C(V_s - V_{c1}) + S_D(V_s - V_o) \tag{13}$$

$$(S_A + S_B + S_C + S_D)L\frac{dI}{dt} = (S_A + S_B + S_C + S_D)V_s + S_B(-V_{c2}) + S_C(-V_{c1}) + S_D(-V_o) \tag{14}$$

Given that $S_A + S_B + S_C + S_D = 1$ and $V_{c1} + V_{c1} = V_o$, (13) and (14) simplifies to (15) and (16)

$$L\frac{dI}{dt} = V_s + S_B(-V_{c2}) + S_C(-V_{C1}) - S_D * (V_{c2} + V_{c2}) \tag{15}$$

$$L\frac{dI}{dt} = V_s - (V_{c1})(S_B + S_D) - (V_{c2})(S_C + S_D) \tag{16}$$

where $S_B + S_D = S_2S_4 + S_1S_4 = S_4$ and $S_C + S_D = S_1S_3 + S_1S_4 = S_1$.

$$L\frac{dI}{dt} = V_s - S_1V_{c1} - S_4V_{c2} \tag{17}$$

Similarly, applying switching functions to the capacitor equations gives

$$C_1\frac{dV_{c1}}{dt}(S_A + S_B + S_C + S_D) = -(S_A + S_B)I_o + S_C(I_L - I_o) + S_D(I_L - I_o) \tag{18}$$

$$C_1\frac{dV_{c1}}{dt} = -I_o + (S_1S_3 + S_1S_4) * I_L \tag{19}$$

$$C_2\frac{dV_{c2}}{dt} = -I_o + (S_4S_1 + S_2S_4) * I_L \tag{20}$$

Given that $S_3 + S_4 = 1$ and $S_1 + S_2 = 1$, therefore (19) and (20) become

$$C_1\frac{dV_{c1}}{dt} = -I_o + S_1I_L, \quad C_2\frac{dV_{c2}}{dt} = -I_o + S_4I_L \tag{21}$$

As shown in (17) and (21), the average value of the inductor voltage and capacitor current is zero,

$$\frac{dI}{dt} = 0; \quad \frac{dV_{c1}}{dt} = 0; \quad \frac{dV_{c2}}{dt} = 0 \tag{22}$$

Inserting (21) into (17) gives

$$V_s = S_1V_{c1} + S_4V_{c2} \tag{23}$$

similarly, summing (20) and applying averaging technique gives

$$-2I_o + S_1I_L + S_4I_L = 0 \tag{24}$$

Taking the duty ratio of switching functions $S_1$ and $S_4$ as $d_1$ and $d_4$, respectively, then (23) becomes:

$$\frac{I_o}{I_L} = \frac{2}{d_1 + d_4} \tag{25}$$

For ideal operating conditions, the output power $P_o$ equals the input power, $P_{in}$ such that the voltage gain of the converter is derived as in (28)

$$I_oV_o = I_LV_i \tag{26}$$

$$\frac{V_o}{V_i} = \frac{I_L}{I_o} \tag{27}$$

$$\frac{V_o}{V_i} = \frac{2}{d_1 + d_4} \tag{28}$$

### *B. Three Phase Inverter Modeling*

The derivation and analysis of the inverter voltage and current equations are given as follows;

$$\frac{V_{dc}}{2}(2S_{ap} - 1) = LpI_{ag} + rLI_{ag} + V_{ag} + V_{nnm} \tag{29}$$

$$\frac{V_{dc}}{2}(2S_{bp} - 1) = LpI_{bg} + rLI_{bg} + V_{bg} + V_{nnm} \tag{30}$$

$$\frac{V_{dc}}{2}(2S_{cp} - 1) = LpI_{cg} + rLI_{cg} + V_{cg} + V_{nnm} \tag{31}$$

where $p = \frac{d}{dt}$. Combining (28)-(30) gives

$$\frac{V_{dc}}{2}(2S_{abcp} - 1) = LpI_{abcg} + rLI_{abcg} + V_{abcg} + V_{nnm} \tag{32}$$

The general expression for transformation from the stationary $abc$ to rotating $dq$ reference frame is illustrated in (32)

$$F_{qdo} = k(\theta)F_{abc} \tag{33}$$

$$\begin{pmatrix} V_q \\ V_d \\ V_o \end{pmatrix} = k(\theta)\begin{pmatrix} V_{ao} \\ V_{bo} \\ V_{co} \end{pmatrix}; \begin{pmatrix} I_q \\ I_d \\ I_o \end{pmatrix} = k(\theta)\begin{pmatrix} I_{ao} \\ I_{bo} \\ I_{co} \end{pmatrix} \tag{34}$$

$$k(\theta) = \frac{2}{3}\begin{pmatrix} cos\theta & cos(\theta-\theta_{eo}) & cos(\theta+\theta_{eo}) \\ sin\theta & sin(\theta-\theta_{eo}) & sin(\theta+\theta_{eo}) \\ \frac{1}{2} & \frac{1}{2} & \frac{1}{2} \end{pmatrix} \tag{35}$$

The switching function, $S_{ip}$ of the proposed inverter can be expressed as a function of its modulation index, $M_{ip}$ such that

$$M_{ip} = 2S_{ip}+1, \frac{V_{dc}}{2}\begin{pmatrix} k(\theta)M_{ap} \\ k(\theta)M_{bp} \\ k(\theta)M_{cp} \end{pmatrix} = \frac{V_{dc}}{2}\begin{pmatrix} M_q \\ M_d \\ M_o \end{pmatrix} \tag{36}$$

The grid resistance, $r$, inductance, $L$, and $w$ are expressed as:

$$\widehat{r} = \begin{pmatrix} r & 0 & 0 \\ 0 & r & 0 \\ 0 & 0 & r \end{pmatrix}; L = \begin{pmatrix} L & 0 & 0 \\ 0 & L & 0 \\ 0 & 0 & L \end{pmatrix} \tag{37}$$

$$\widehat{\omega} = \begin{pmatrix} \omega & 0 & 0 \\ 0 & -\omega & 0 \\ 0 & 0 & \omega \end{pmatrix} \tag{38}$$

Notice that $\omega = 2\pi f$ where $f$ is the frequency of the grid.

$$\frac{V_{dc}}{2}(M_q) = rI_{qg} + LpI_{qg} - \omega LI_{qg} + V_{qg} \tag{39}$$

$$\frac{V_{dc}}{2}(M_d) = rI_{dg} + LpI_{dg} - \omega LI_{dg} + V_{dg} \tag{40}$$

Further, the inverter input current, $I_o$, is given by the expression

$$C_pV_{pv} = I_{pv} - I_o, \quad I_o = \frac{3}{4}(M_qI_q + M_dI_d) \tag{41}$$

$$P = \frac{3}{2}(V_{qg}I_{qg} + V_{dg}I_{dg}), \quad Q = \frac{3}{2}(-V_{qg}I_{dg} + V_{dg}I_{qg}) \tag{42}$$

Moreover, the active power, $P$, and reactive power, $Q$ in (40) are related by the expression

$$Q = Ptan\theta_{eo} \tag{43}$$

Inserting (40) in (41) and aligning the $q$-axis and $d$-axis grid voltages such that the $V_{dg}$ = 0 and $\theta_{eo} = \theta$, then (40) simplifies to (42)

$$P = \frac{3}{2}(V_{qg}I_{qg}); Q = \frac{3}{2}(-V_{qg}I_{dg}); I_{dg} = I_{qg}tan\theta \tag{44}$$

where $\theta = cos^{-}1(pF)$, $pF$ is the grid power factor, $I_{dg}$ and $I_{qg}$ are the $d$-axis and $q$-axis grid current, respectively. At steady state, $pI_{qg} = 0$; $pI_{dg} = 0$. Therefore, the corresponding steady state expression for $M_q$ and $M_d$ are depicted below

$$M_q = \frac{2}{V_o}(rI_{qg} + \omega LI_{qg}tan\theta + V_{qg}) \tag{45}$$

$$M_d = \frac{2}{V_o}(rI_{dg}tan\theta - \omega LI_{dg}tan\theta + V_{qg}) \tag{46}$$

TABLE II
PARAMETER SPECIFICATION

| Parameter | Symbol | Unit | Value |
|---|---|---|---|
| Line Inductance | $L$ | $mH$ | 0.2 |
| Line Capacitance | $C$ | $F$ | 250 |
| Line resistance | $r$ | $\Omega$ | 0.1 |
| Switching frequency | $f_{sw}$ | $kHz$ | 18 |
| Grid Voltage | $V_g$ | - | 160 |
| Input voltage | $V_s$ | $Volt$ | 400 |
| Target power | $P_{out}$ | $Watt$ | 600 |

## III. MODEL SIMULATION

Detailed mathematical model of the dynamic behavior of the proposed grid connected fuel cell distributed energy system has been presented in section I. In addition, steady state expression of the TLC buck boost converter and three phase grid connected inverter is also presented. Based on the obtained steady state voltage gain of the three-level bidirectional buck boost converter, it is apparent that the proposed system does have the flexibility to operate in either buck and boost mode subject to the value of the switching signal or duty cycle. Using the dynamic of the fuel cell, a programming script was written in MATLAB to evaluate the values of $M_q$, $M_d$, $I_{dg}$ and $I_{qg}$. The model simulation parameters are specified as in Table II. Based on the respective values of the given leading and lagging $pF$, the derived steady state equations are used to evaluate the values of $M_q$, $M_d$, $I_{dg}$ and $I_{qg}$. Given the parameterized values of lagging and leading $pF$ as 0.2, 0.4, 0.6, 0.8 and -0.2, -0.4, -0.6, -0.8, respectively, the corresponding values of $M_q$, $M_d$, $I_{dg}$, $I_{qg}$ and grid active power, $P_g$ are disclosed in Table III and Table Table IV, respectively.

TABLE III
$M_q$, $M_d$, $I_{dg}$ AND $I_{qg}$ SIMULATIONS RESULTS FOR LAGGING $pF$

| $pF$ | $M_q$ | $M_d$ | $I_q(A)$ | $I_d(A)$ | $I_{peak}(A)$ | $P_g(W)$ |
|---|---|---|---|---|---|---|
| 0.2 | 0.779 | −0.010 | 2.41 | −11.8 | 12.04 | 598 |
| 0.4 | 0.791 | −0.008 | 2.47 | −5.7 | 6.91 | 594 |
| 0.6 | 0.795 | −0.006 | 2.49 | −3.3 | 4.15 | 597 |
| 0.8 | 0.795 | −0.006 | 2.5 | −1.9 | 3.12 | 598 |

TABLE IV
$M_q$, $M_d$, $I_{dg}$ AND $I_{qg}$ SIMULATIONS RESULTS FOR LEADING $pF$

| $pF$ | $M_q$ | $M_d$ | $I_q(A)$ | $I_d(A)$ | $I_{peak}(A)$ | $P_g(W)$ |
|---|---|---|---|---|---|---|
| −0.2 | 0.779 | −0.010 | 2.41 | −11.8 | 12.04 | 598 |
| −0.4 | 0.791 | −0.008 | 2.47 | −5.7 | 6.91 | 594 |
| −0.6 | 0.795 | −0.006 | 2.49 | −3.3 | 4.15 | 597 |
| −0.8 | 0.795 | −0.006 | 2.5 | −1.9 | 3.12 | 598 |

In order to determine the active power injection for leading and lagging $pF(s)$, the complete system structure was modeled in PLECS software.

## IV. RESULTS AND DISCUSSION

### A. Model Simulation Results and Analysis

The simulation waveforms for $M_q$ vs. $M_d$, grid active power, $P_g$ vs. $pF$ and $I_g$ vs. $pF$ are presented in this section. In

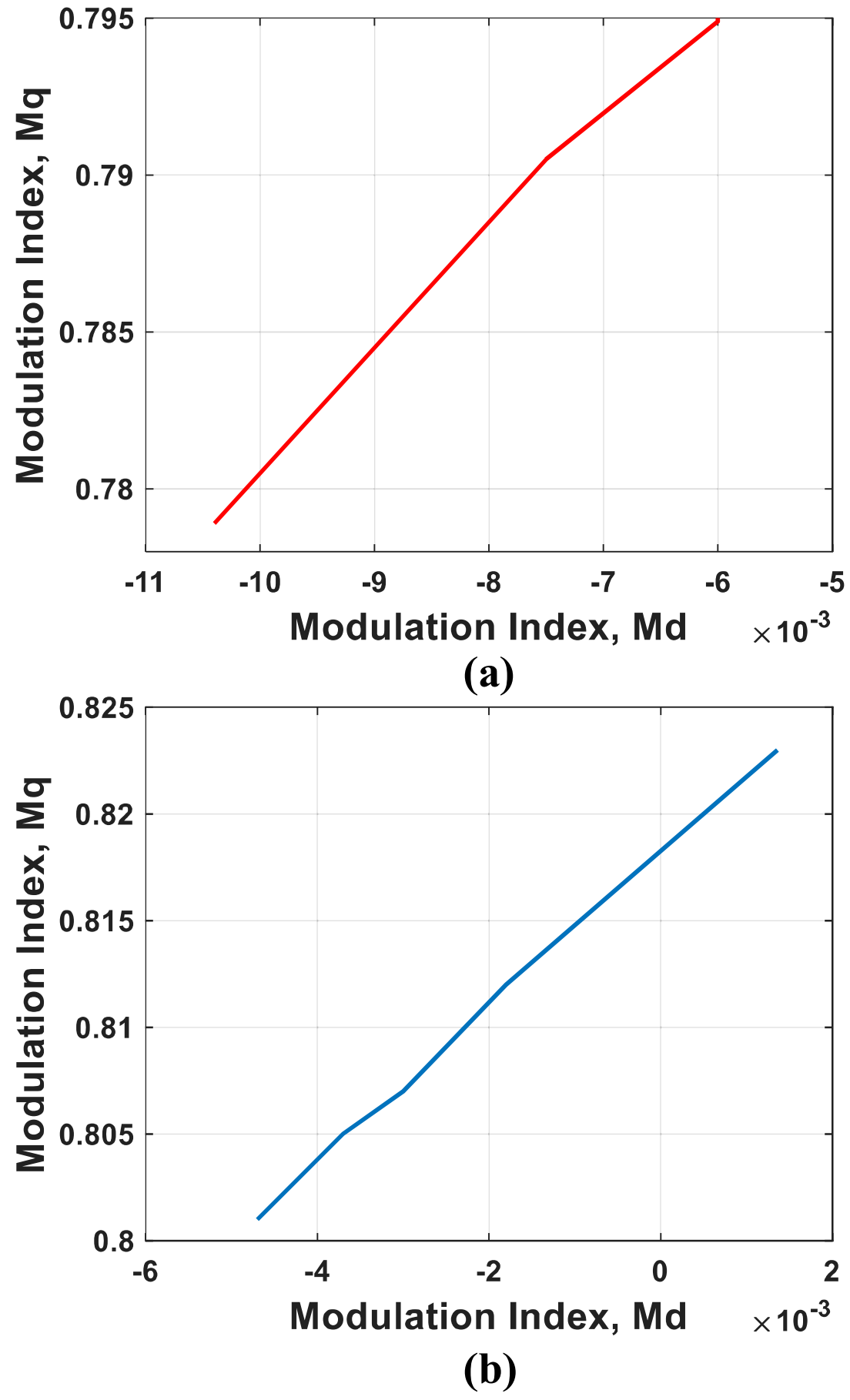

Fig. 6. Plot of $M_q$ vs. $M_d$ for (a) lagging power factor (b) leading power factor

.

addition, leading and lagging power factor values ranging from 0.2 to 0.8 with an offset of 0.2 is harnessed for the simulation. The plot of $d-axis$ modulation index against $q-axis$ modulation index, ($M_q$vs.$M_d$) for lagging and leading $pF$ are utilized for the simulation in Fig.6(a) and (b), respectively. For both leading and lagging power factors scenarios, it is observed that the wave-forms assume a straight line implying a linear relationship between both indices, with the maximum value of $M_q$ for lagging power factors being 0.825 while the minimum power factor is obtained as 0.002. Moreover, the waveform of grid current magnitude $I_{qd}$ versus lagging and leading $pF$ are shown in Fig. 7(a) and 7(b), respectively. It is observed that increasing the values of the power factors results in higher current injection to the grid, in which case the maximum grid current corresponds to unity power factor. At this juncture, both the grid current and voltage are fully in phase, leading to maximum power delivery to the grid. As shown in Fig. 8(a) and Fig. 8(b), the effect of power factor correction on the grid is such that an increase in the power factor culminates in a corresponding increase in the current and active power injection to the grid and vice versa. Moreover, the effect of lagging and leading power factors in this grid connected inverter is mainly due to the presence or penetration of capacitive and inductive components.

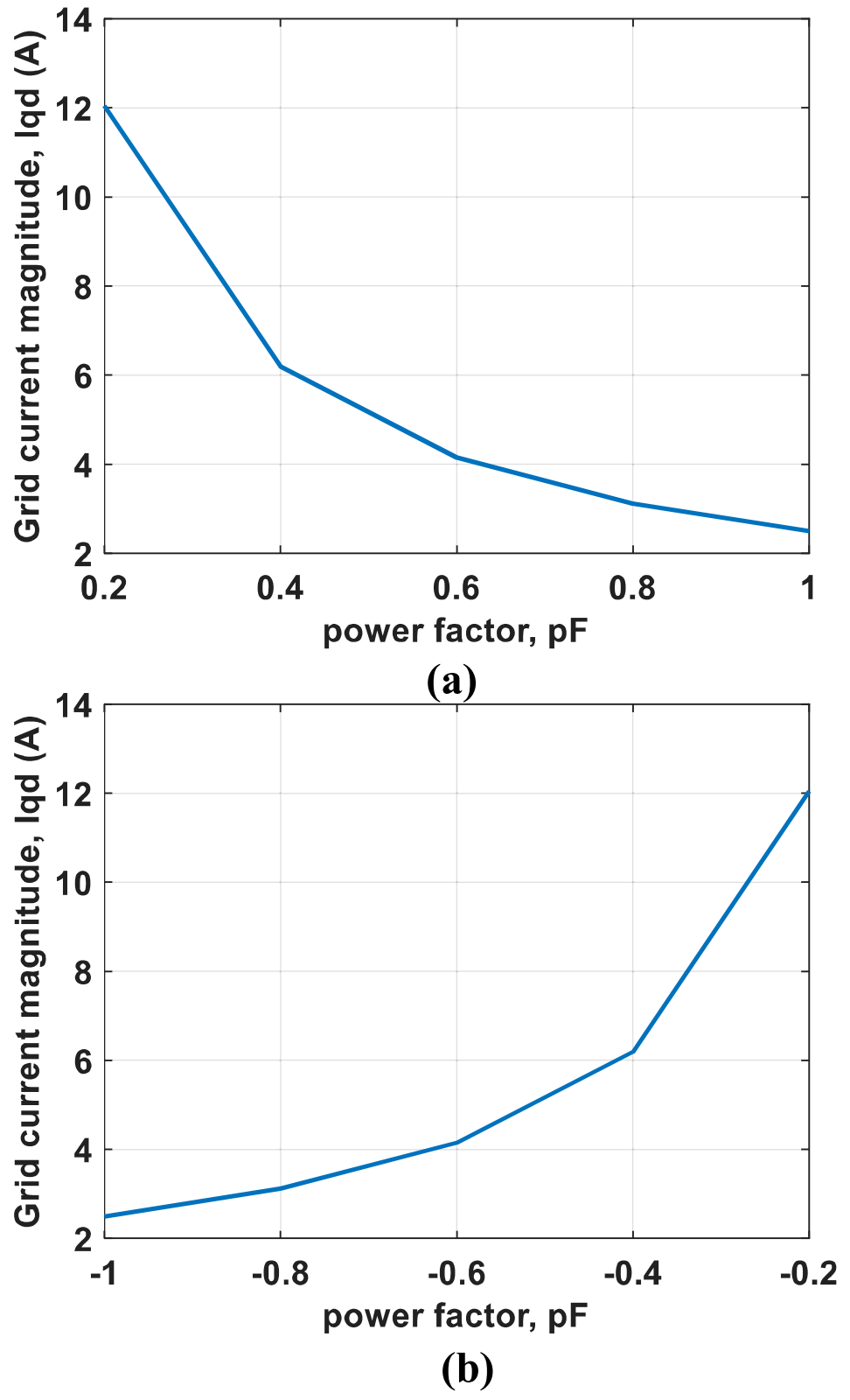

Fig. 7. Plot of $I_{qd}$ vs. $pF$ for (a) lagging power factor (b) leading power factor

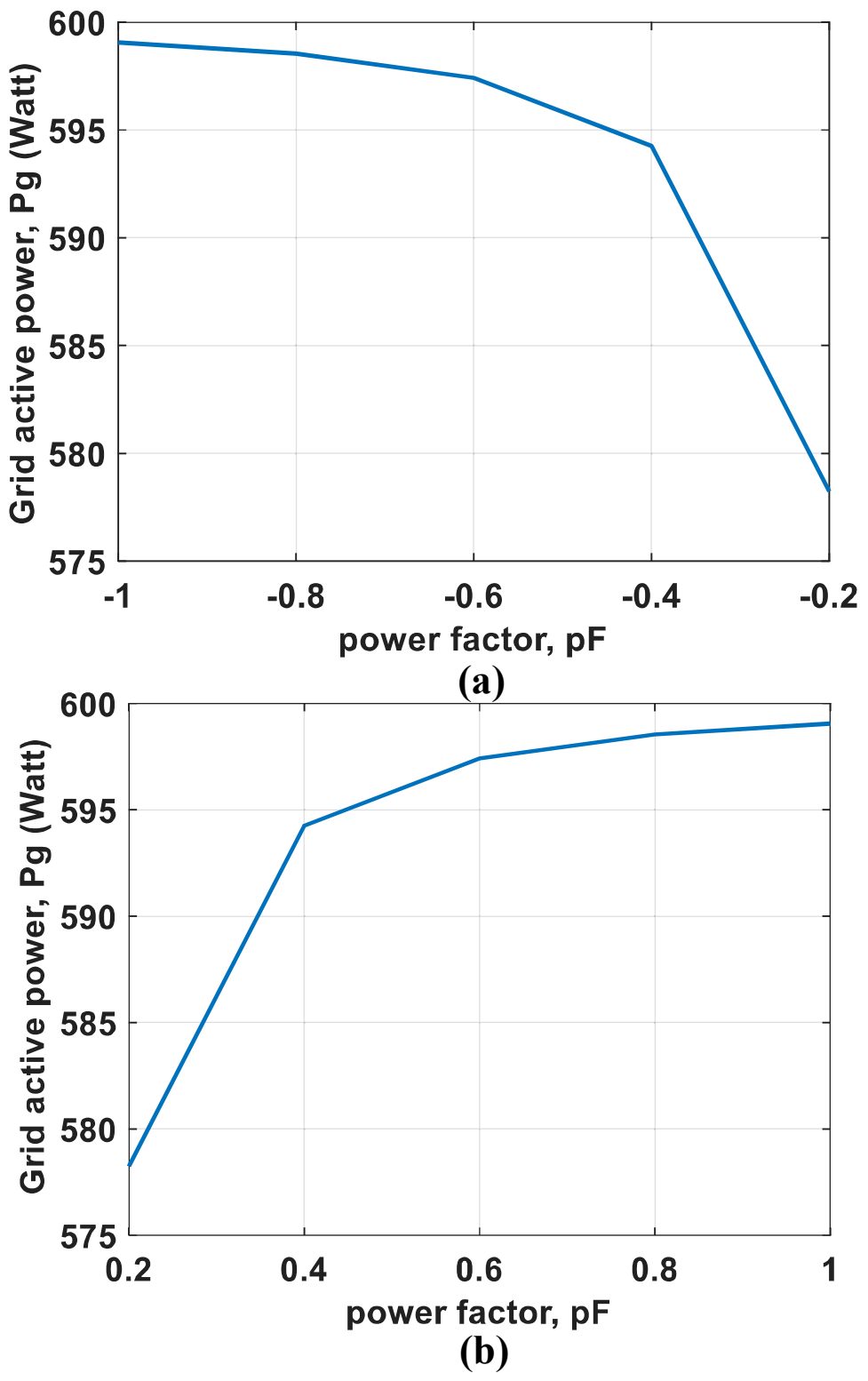

Fig. 8. Plot of grid power,$P_g$ vs. power factor, $pF$ for leading power factor

.

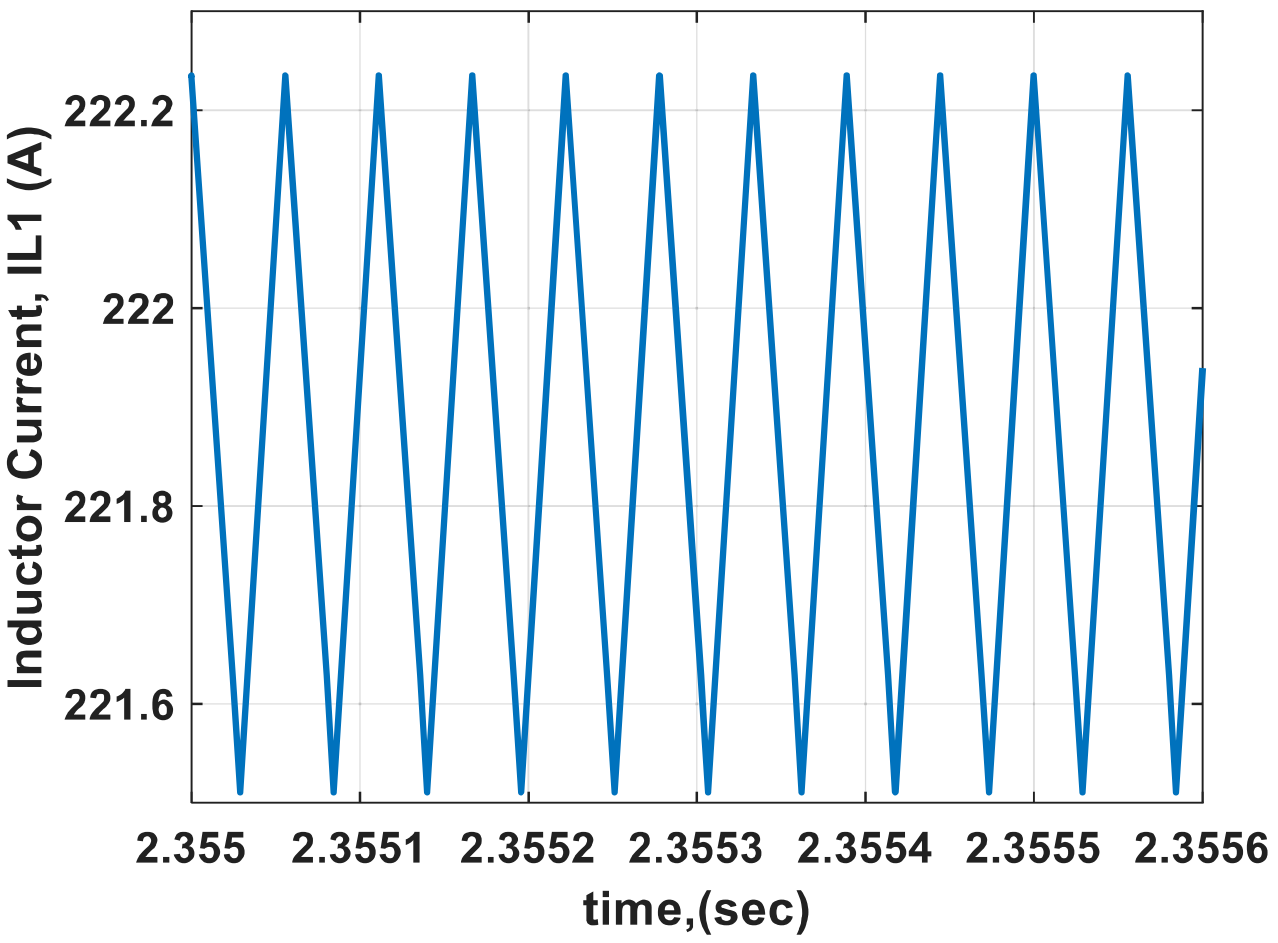

Fig. 9. Inductor current waveform, $I_{L1}$

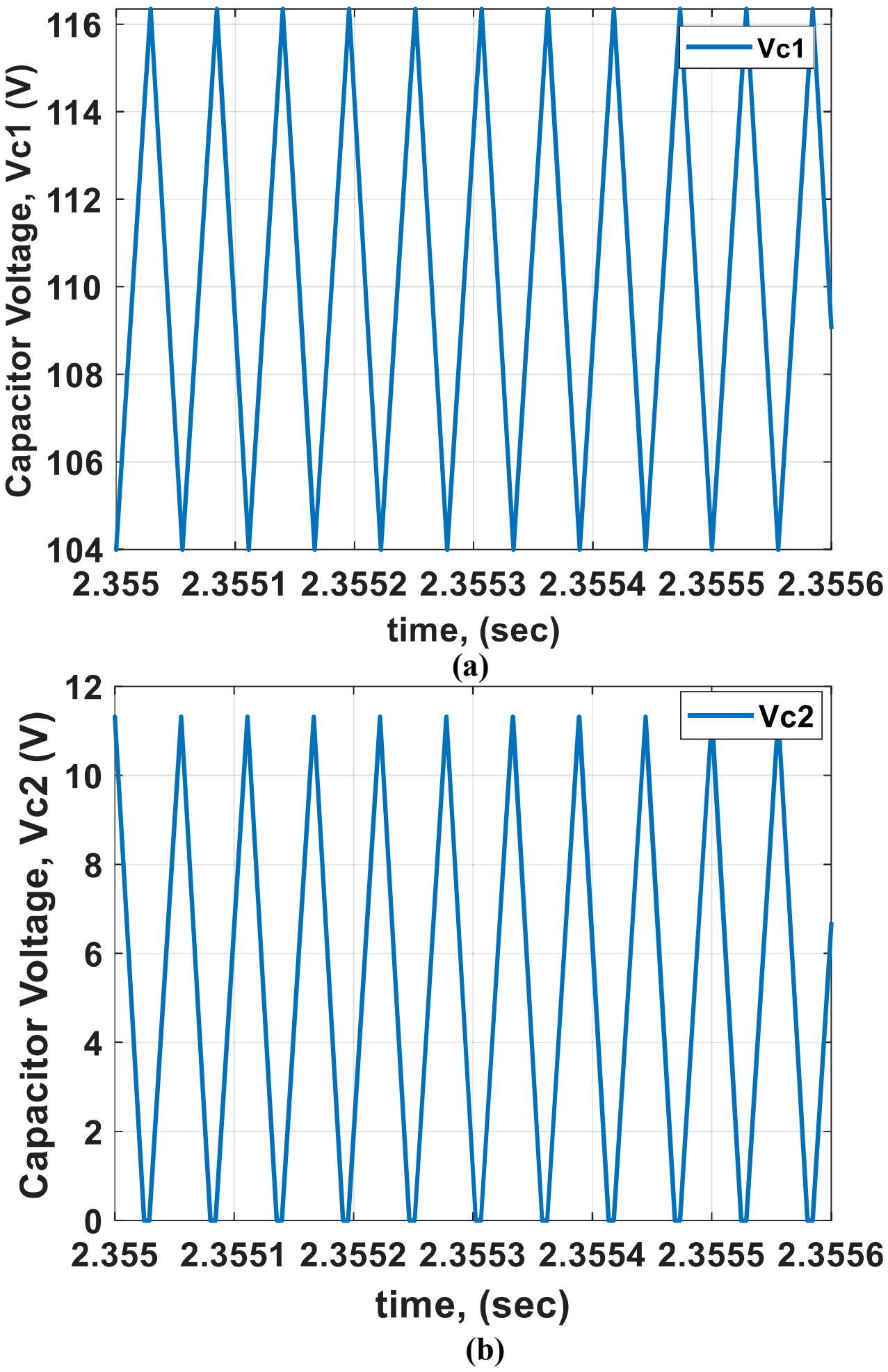

Fig. 10. Capacitor Voltage Waveform, (a) $V_{c2}$ (b) $V_{c1}$

The complete system set-up comprising the ultra-capacitor battery bank/fuel cell, the DC-DC three level bidirectional buck boost converter and a three phase DC-AC converter are modeled in PLECS simulation environment. In addition, the model simulation parameters are based on parameter values in table I coupled with the numerically computed modulation indices, $M_q$, and $M_d$ as contained in Table II and Table III. Meanwhile, the simulation waveforms for the converter side inductor current, $I_{L1}$ and capacitor voltages ($V_{c1}$ and $V_{c2}$F) are shown in Fig. 9 and Fig. 10, respectively. Apparently, for a grid phase angle of $\frac{\pi}{36}$ rad, 160V grid voltage and considering a simulation time of $2ms$, the complete model simulation shows that a corresponding grid current of $220A$, and an active power, $P_g = 5 \times 10^4$W are injected to the grid. Both the grid current and $P_g$ waveform are denoted by Fig. 11 and Fig.12, respectively.

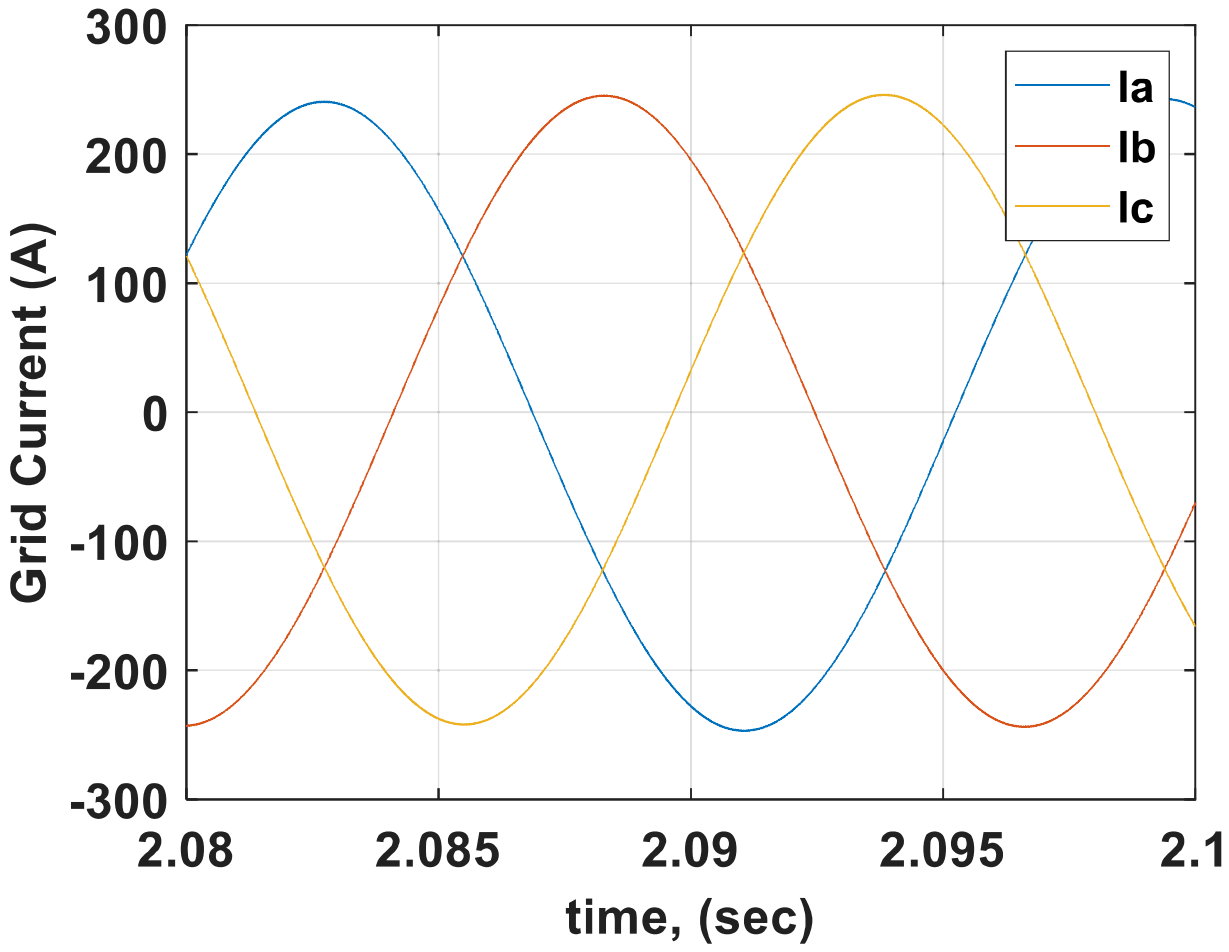

Fig. 11. Plot of grid current against simulation time

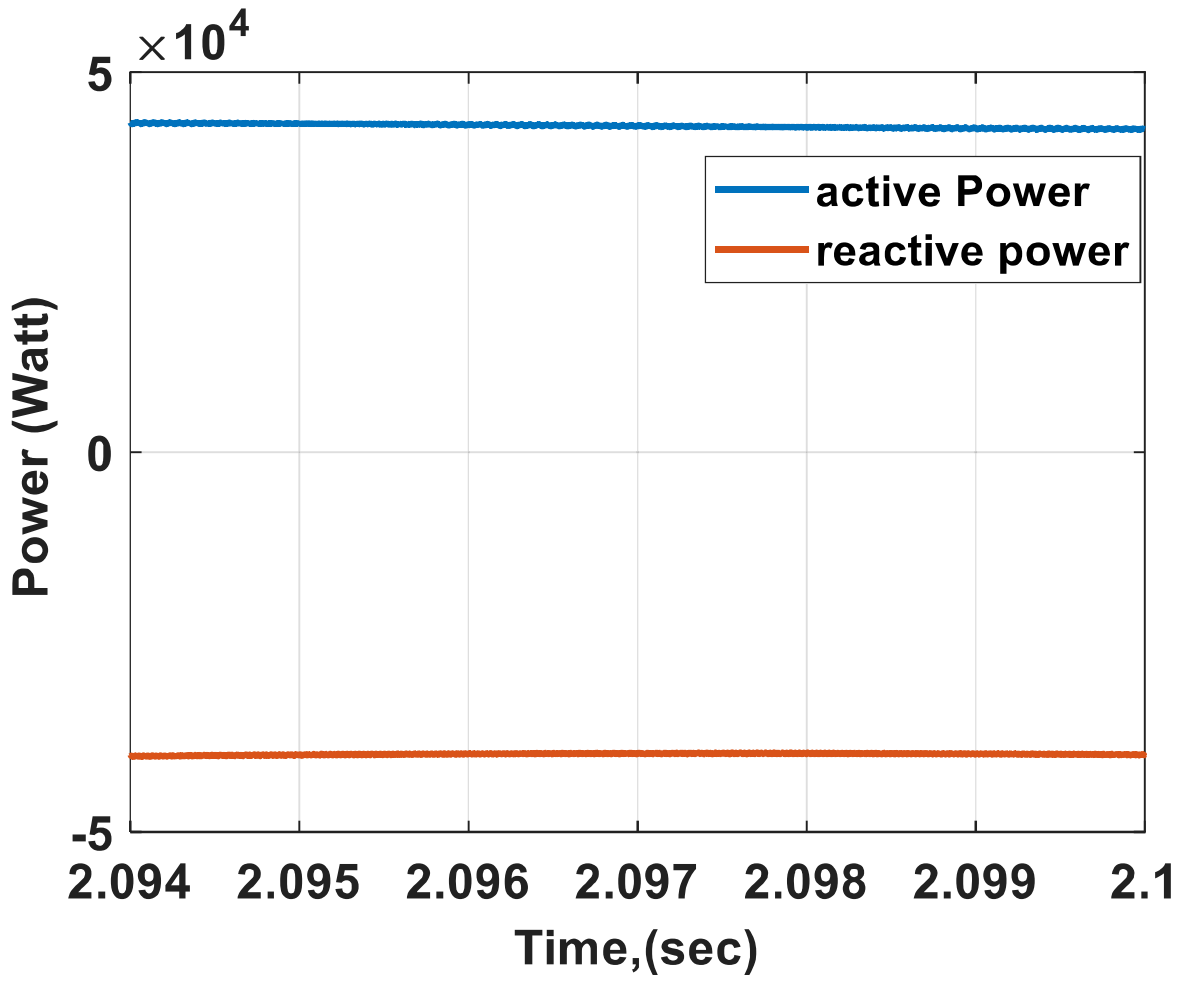

Fig. 12. Plot of active power and reactive power versus time

## V. CONCLUSION

The model derivation of a three-phase bidirectional inverter has been analyzed and modeled. In addition, the grid connected three phase inverter has also been modeled, mathematically. Further, the performance of both converters under steady state and dynamic conditions have been simulated while offering a comprehensive discussion of the corresponding simulation waveforms. The modulation indices were obtained from the evaluation of model equations based on the specified parameters. The behavior of the grid is such that at non-unity power factor, power is been absorbed from the grid resulting in losses and reduced efficiency of the system. On one hand, the three-phase bidirectional converter is useful for voltage regulation depending on the desired voltage level of the connected device while on the other hand, its bidirectional capability enhances a dual level power flow from the fuel cell to grid as well as from the grid to fuel cell for charging operation in the event of grid outage